# Computing at Hasylab: Perl/PerlTk is the new scripting language for Spectra

Th. Kracht

*Hamburger Synchrotronstrahlungslabor HASYLAB at Deutsches Elektronen-Synchrotron DESY, Notkestrasse 85, 22603 Hamburg, Germany*

**Abstract**

At the Hamburger Synchrotronstrahlungslabor HASYLAB most experiments are controlled by the online program Spectra, which runs on Linux PCs. Spectra has been converted into a Perl module, allowing online programmers to write their code in Perl. They can take advantage of complex data structures, flow control, subroutines, OO features, the network interface, etc. In addition, Perl offers other features that are well suited to make the online functionality of Spectra (monochromator and diffractometer routines, etc.) available to the programmer.

## 1. Introduction

At the Hamburger Synchrotronstrahlungslabor HASYLAB the program Spectra [1],[2] is used for data acquisition and instrument control purposes at most of the beamlines. It runs on Linux PCs and accesses the electronics via VME, Camac, GPIB, Can and RS232. Except for the serial lines, which are driven by terminal servers, all other busses are connected by PCI or AT bus interfaces (fig. 1).

Spectra is the standard software tool at HASYLAB. It is flexible enough to meet the requirements of many different experimental stations. This has been accomplished by allowing the specification of measurements in several ways:

1. Most of the measurements are energy, diffractometer or single motor scans. They are performed with the help of menus. The details of the measurements are specified in scripts which are executed at the beginning, during and at the end of the scans.

2. It is possible to execute a series of scans from scripts using the same scan function that is invoked by the menus.

3. If a special measurement does not fit into the scheme of the standard scan routine, the whole procedure has to be coded explicitly in a script.

Each of these methods relies on the internal Spectra interpreter. It consists of a parser that analyzes the command line and a set of functions that are able to execute control commands, which change the flow of program execution. Typical control commands are: for, if-else, goto, gosub. Notice that the interpreter is independent of any online functionality. The same applies to most of the Spectra command verbs and functions. They provide functionality that is heavily used in scripts: symbols, mathematical functions, string handling functions, a socket interface, I/O functions, etc.

The syntax of the original Spectra command language imitates a language that was frequently used at the time Spectra was launched. It is considerably different from what a shell programmer expects now.

To conclude: one reason for modernizing the Spectra scripting language is the acceptance by the intended users. Another reason is the reduction of the software maintenance effort: in the future Spectra should concentrate of the online functionality only, employing a generic scripting language for the rest of the job.

## 2. Reasons for selecting Perl

It is certainly not possible to describe the language Perl in a few sentences. In the following, only the most important items are discussed which make it the best candidate for being the Spectra scripting language:

1. Perl is apropriate for our purpose. It has bit manipulation operators, which are mandatory when we have to access hardware registers, and at the same time it offers an excellent interface to the operation systems, which is well suited for directory or file operations. The way how lists and hashes are implemented is very elegant. Regular expressions are useful when we have to deal with strings.

2. Perl is widely distributed and appears to dominate the shelves in the bookstores. It is frequently used for CGI programming and system administration. Therefore, those who know Perl from other applications have an easy access to Spectra and those who learn Perl to run Spectra can take advantage of this knowledge.

3. Perl being an interpreter reduces the code-and-test cylce time compared to other applications, which are written in a non-interpreted language.

4. Perl is an extendable language. Many useful modules are are freely available in the web.

5. PerlTk is a module that interfaces Perl to the widget library Tk.

## 3. The Spectra – Perl interface

An online session begins by starting Spectra. The program initializes the hardware and opens log files. One log file keeps track of motor positions; another file stores user commands and scan parameters. It would be very inefficient to go through this procedure for every Perl script that is executed. Thus, it must be possible to call scripts from the Spectra command line leaving Spectra active while the script executes (Of course scripts may call other scripts). The interface has been implemented using Perl library functions and a piece of Perl code that installs a persistent interpreter. This means that the interpreter is loaded into the memory only once per session.

The second part of the implementation makes the Spectra functionality available for Perl. A module, Spectra.pm, has been coded that connects Perl and Spectra with the help of the XS interface [3]. The detailed description of Spectra.pm can be found in the web [4]. Here we discuss some key features only, in particular the interfaces to

1. <u>Spectra symbols</u>. They are defined and translated using the %SYM hash, e.g.:

   ```
   # define the symbol d_crystal
   $SYM{d_crystal} = 5.4309;

   # print the current value of d_crystal
   print " d_crystal = $SYM{d_crystal}\n";
   ```

   A hash is a list that is indexed by strings. The %SYM hash makes use of the Perl data wrapping mechanism. It is tied to a package that translates all accesses to %SYM into function calls to Spectra.

2. <u>Graphical queue elements (GQE)</u>. GQEs are internal data structures. Each of these is described in a class. The most important being SCAN, which refers to a structure that consists of an info block and two columns of floating point numbers (x, y). The x-component typically stores an energy or a motor position, the y-component some measured quantity. The following example shows how a SCAN can be created:

   ```
   $s1 = SCAN->create( name => "Ge",
                       title => "A Title",
                       np => 1000,
                       xlabel => "x-Descr.",
                       ylabel => "y-Descr.");
   ```

   Several methods are defined for such an object, e.g.:

   ```
   $col_org = $s1->get( colour); # get the current colour
   $s1->configure( colour => 2); # red
   $s1->display();               # display the scan
   ```

   The data wrapper makes the data columns accessible. The following example shows how the value of the first data point is changed:

   ```
   $s1->{x}[0] = 0;
   $s1->{y}[0] = 1;
   ```

3. The monochromator and diffractometer system is presented to the user by functions and the %EXP hash. Both methods appear in the following piece of code that demonstrates how the energy is changed and read:

   ```
   # Change/read the energy …
   # … by function calls
   Energy(8980) or die "Failed to set the energy";
   $ener = Energy();
   print "Current energy $ener\n";

   # … and via the %EXP hash
   $EXP{energy} = 8980;
   print " Current energy $EXP{energy}\n";
   ```

   The interface to the diffractometer software is very similar. The Q() function sets or returns the entire q-vector. In addition, the components of the q-vector can be set separately.

   ```
   # set/get the complete q-vector
   Q( qx => 1, qy => 0, qz => 0, psi => 0) or die "Failed to set Q ";
   ($qx, $qy, $qz, $psi) = split ' ',Q();

   # … or a single component.
   $ret = Qx(1);
   $qx = Qx();

   # the %EXP hash provides the same functionality
   $EXP{q} = "$qx $qy $qz $psi";
   ($qx, $qy, $qz, $psi) = split ' ', $EXP{q};
   $EXP{qx} = 1;
   $qx = $EXP{qx};
   ```

4. <u>Stepping motors</u> are available as objects:

   ```
   # create an instance of the Motor class
   $m11 = Motor->locate( name => "mot11");

   # configure() changes parameters
   $m11->configure(slew => "10000",
                   base => "100");

   # read the current position
   $pos = $m11->get(position);

   # move the motor to a position
   $m11->configure(position => $pos);
   ```

   Besides this object-oriented approach, stepping motors may be operated by simple function calls. This becomes necessary, if a set of motors, which does not belong to a monochromator or diffractometer system, has to be moved simultaneously

   ```
   Move( mot11 => $pos11,
         mot12 => $pos12)
      or die "Failed to move mot11 and mot12 ";
   ```

5. <u>Low level I/O routines.</u>

   Although low-level I/O is rarely used, the Spectra-Perl interface would be incomplete, if functions that address the data busses VME, Can and GPIB directly would be missing.

   ```
   # VME
   $ret = Vme( A24D16, $base, $offset);         # read
   $ret = Vme( A24D16, $base, $offset, $datum); # write

   # Can bus
   ($adr, $len, @msg) = split ' ',Can_read();
   $ret = Can_write( 0x201, 0, 0);

   # GPIB
   $buffer = Gpib_read( $id, $len_max);
   $ret = Gpib_write( $id, $msg);
   ```

These examples show how the online functionality of Spectra is ported to Perl using functions, objects and the data wrapper. Perl proved to be a proper language this kind of applications.

## 4. Graphical user interfaces with PerlTk

The PerlTk module is an extension to Perl which provides a complete widget library, including label, entry, listbox, dialogbox, scale, button, menu and text widgets. The library is sufficient to create the Spectra graphical user interface (GUI). It is planned to provide GUIs for all applications that are presented as menus so far. In addition there will be the option to include GUIs for specific purposes. Notice that since the GUI code is interpreted, extensions can be easily included.

## 5. Conclusions

Perl became the new Spectra scripting language at Hasylab. It has been demonstrated that it is very well suited for online purposes. The interface is in operation at different beamlines. The original Spectra scripting language is still supported. It is planned to facilitate PerlTk for the development of the Spectra GUI. Partially the GUI has already been ported.

**References**

[1] Th. Kracht, SPECTRA, A Program Package for the Analysis and Presentation of Data, Version March 18, 2002, http://www-hasylab.desy.de/services/computing/spectra/spectra.html

[2] Th. Kracht, ONLINE, A Program Package for Data Acquisition and Beamline Control at Hasylab, Version October 1, 2002, http://www-hasylab.desy.de/services/computing/online/online.html

[3] perlxs, http://www.perldoc.com/perl5.6/pod/perlxs.html

[4] Th. Kracht, Perl-Spectra, http://www-hasylab.desy.de/services/computing/perl_spectra/perl_spectra.html

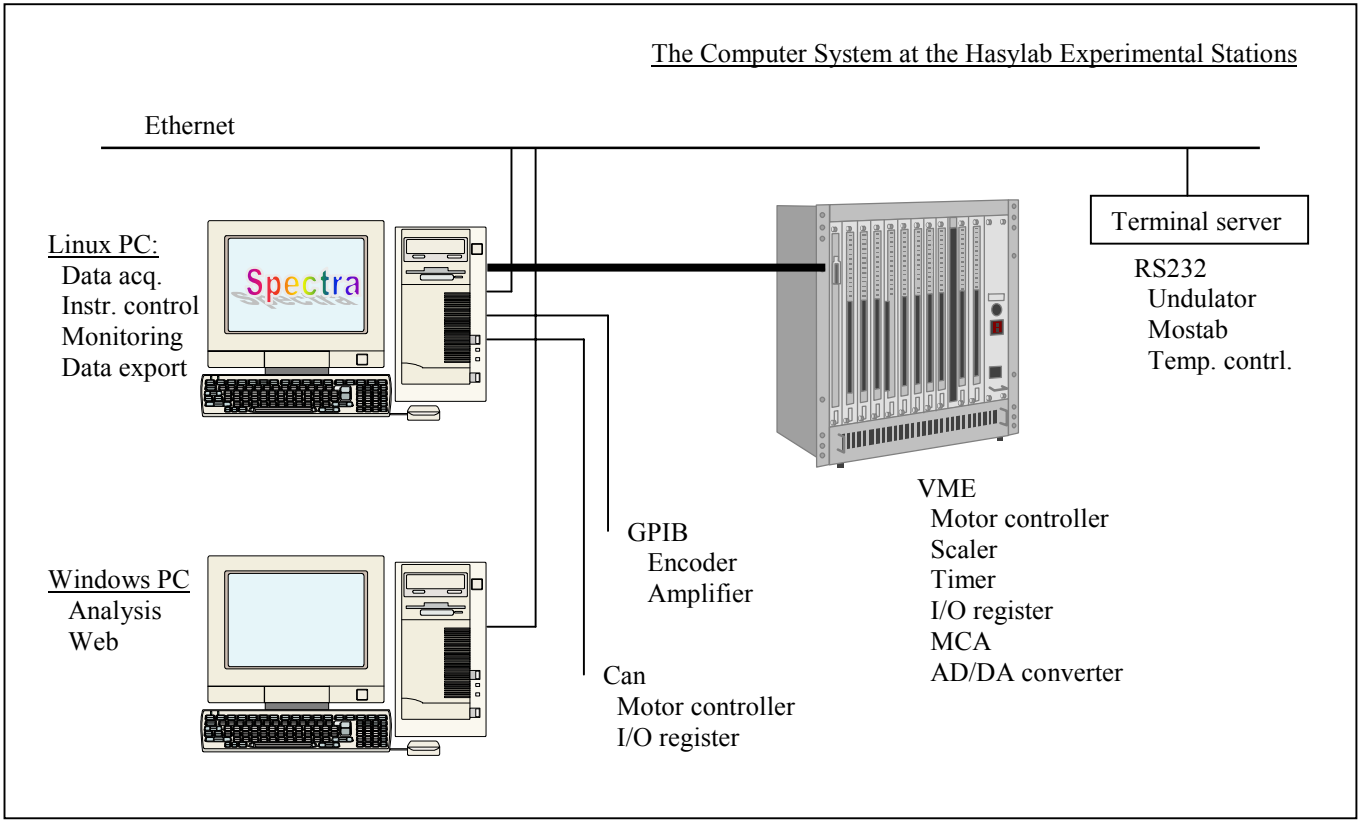

Figure 1

**Figure captions**

Fig. 1: The computer system at the Hasylab experimental stations